# A Range Matching CAM for Hierarchical Defect Tolerance Technique in NRAM Structures

Hossein Pourmeidani
Computer Engineering Department
Azad University of Arak
Isfahan, Iran
hpooor@gmail.com

Mehdi Habibi
Electrical Engineering Department
Isfahan University
Isfahan, Iran
mhabibi@eng.ui.ac.ir

*Abstract*— Due to the small size of nanoscale devices, they are highly prone to process disturbances which results in manufacturing defects. Some of the defects are randomly distributed throughout the nanodevice layer. Other disturbances tend to be local and lead to cluster defects caused by factors such as layer misintegration and line width variations. In this paper, we propose a method for identifying cluster defects from random ones. The motivation is to repair the cluster defects using rectangular ranges in a range matching content-addressable memory (RM-CAM) and random defects using triple-modular redundancy (TMR). It is believed a combination of these two approaches is more effective for repairing defects at high error rate with less resource. With the proposed fault repairing technique, defect recovery results are examined for different fault distribution scenarios. Also the mapping circuit structure required for two conceptual 32×32 and 64×64 bit RAMs are presented and their speed, power and transistor count are reported.

I. INTRODUCTION

With complementary metal oxide semiconductor (CMOS) technology reaching the scaling limits, the need for alternative technologies has become necessary. Nanotechnology based fabrication is expected to offer the extra density and potential performance to take electronic circuits to the next step. Nanometer technology promises dramatic increases in device density, but reliability is decreased as a side-effect. Nanometer technology will have smaller, faster transistors, but greater sensitivity to defects such as copper voids, lattice dislocations, parasitic leakage etc and hence, defect tolerance techniques are necessary [1-3].

In the context of reliable nanoelectronics, two main approaches have been proposed: defect tolerance and defect avoidance [4]. Defect tolerance techniques are based on adding redundancy in the design to tolerate defects or faults. Recently, fault tolerance techniques such as triple-modular redundancy (TMR), History Index [5], $N^2$ transistor structure [6] and Logic Code Division Multiple Access (LCDMA) [7] have been investigated. Almost all redundancy based strategies rely on a majority voting. The voter, therefore, becomes a critical unit for the correct operation of any NMR system. In [7], a voter less fault-tolerant strategy to implement a robust NMR system design was proposed. The authors show that using a novel fault-tolerant communication mechanism, namely LCDMA, can transfer data with extremely low error rates among modules and completely eliminate the need for a centralized voter unit.

Defect avoidance techniques however are based on identifying the defects by using an external tester and avoiding them possibly through the use of reconfigurable blocks. Several successful defect avoidance methods have been demonstrated by researchers, some of the most promising being HFTM system [8], tagged repair techniques [9] and BIRA with two-level redundancy [10].

It is expected, however, that nanodevices will suffer from significantly increased permanent failure rate mainly because of the fundamental limitations of the fabrication processes that limit the yield of such devices. The method used in this work has been derived from the combination of defect tolerance and avoidance approaches.

The rest of the paper is organized as follows; the previously reported RM-CAM and TMR fault recovery methods and also the proposed combinational repairing technique are described in Section 2. Experimental results are presented in Section 3. Finally Section 4 presents some concluding remarks.

II. PROPOSED REPAIR TECHNIQUE

A. RM-CAM Based Scheme

In this method the goal is to find rectangular regions which cover a large number of defects without covering a large amount of usable memory. In Fig. 1, a sample distribution of faults with cluster defects has been shown. The following algorithm starts with one mask with small dimensions that moves through the defect matrix achieved from the external tester. When the mask is placed at different positions, the number of defects must be counted. The mask with maximum defects is defined as maximum mask. A threshold is defined for the number of defects per mask. The appropriate threshold value and initial mask dimensions are obtained by considering the limited data entries available on the RM-CAM. Since each cluster requires a predetermined number of entries in the RM-CAM, thus clusters with smaller size than primary mask size and lower density than the threshold, are repaired later with the TMR method. If the maximum mask has defects more than the threshold value it is understood that a cluster is located at that position. To obtain the final mask size that covers the entire cluster, the mask is stretched until nearly all the cluster defects are covered by the mask. Since defect density decreases at the boundaries of the clusters, the mask is stretched from the four sides until the number of defects on the row and column boundary reaches a predetermined value. The defects inside the mask are removed from the initial fault matrix and the process is repeated until no maximum mask with defects higher than the threshold is found and all clusters are identified. The following pseudo code shows the cluster defect identification procedure.

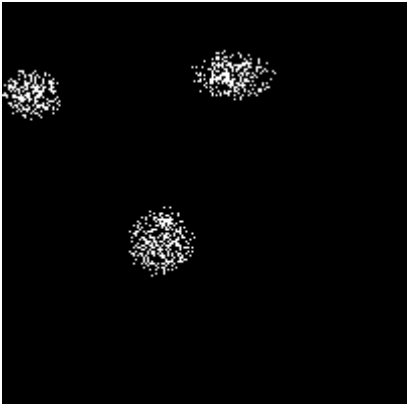

Figure 1. example of cluster defects

**Algorithm 1** finding clusters

1: **while** defects number of maximum mask >= threshold **do**
2:   Finding mask with maximum defects
3:   Stretch mask from four sides until number of defects on
     boundary rows and columns reach the threshold point.
4: **end while**

A searching problem may typically involve finding if a query point is contained in agiven set of ranges. A CAM is a memory in which data is searched with respect to the content stored in the memory. Previously, CAM entries have been used to store the location of defects [8]. In this paper, we present a range searching circuit structure using CAMs. For a given input, a CAM compares it against all of the entries in parallel, and returns the entry that matches the input. An entry value matches the input if the input and the entry value are identical. The rang matching CAM (RM-CAM) structure presented here returns the entry which resides between the lower and upper limit of the input. Fig. 2 shows the lower and upper RM-CAM cell. Unlike the original CAM where a match signal is produced when the input entry is equal to the stored value, the upper and lower RM-CAM blocks check whether the given input is greater or less than the stored value. The READ and WRITE operations are similar to an SRAM cell in CAM. But the searching operation is implemented using XNOR, NAND and AND logic. Gate 3 implement the XNOR, gate 1 and 2 the AND and Transistor N1-N3 the NAND logic as shown in Fig 2. In Fig 2(a) Result is "1" if a given input is greater than the stored and in Fig 2(b) when the given input is less. Signal $P_{out}$ is activated when an equal state happens and is used as the propagate line. The data to be searched is placed on the A and it is compared to the entries stored in the B. The input pulse is applied at the Result. When the input pulse is high, Result is precharged to $V_{dd}$.

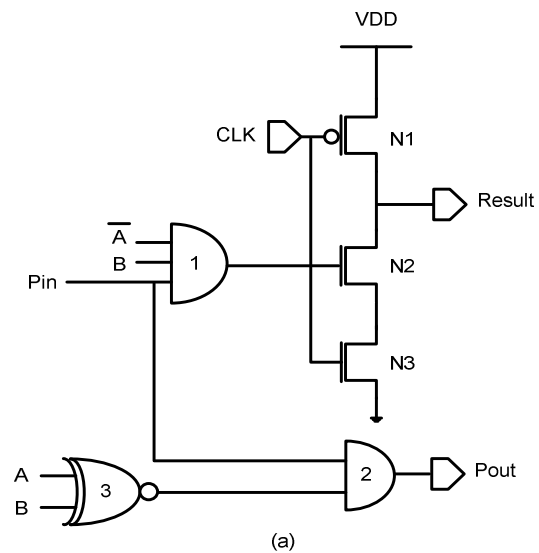

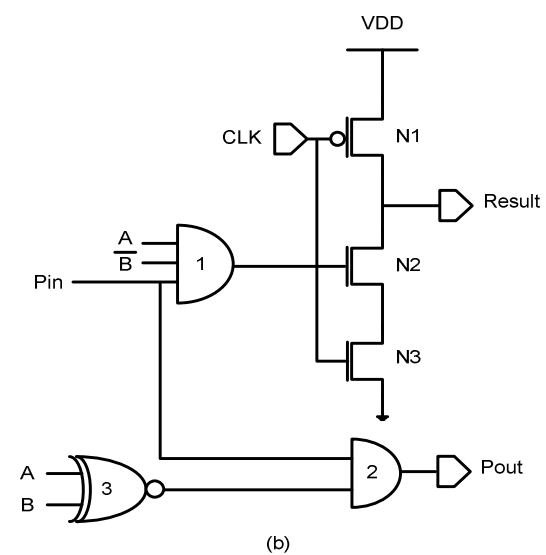

Figure 2. (a) lower CAM cell (b) upper CAM cell

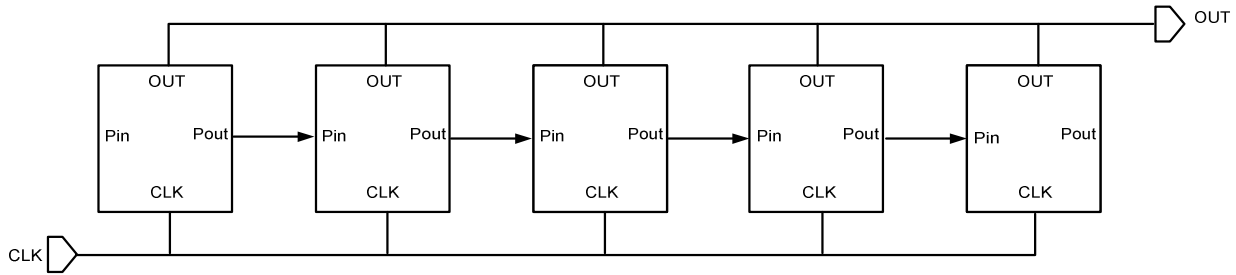

Figure 3. one entry of RM-CAM

In Fig.3 a one entry for RM-CAM is shown. Output line OUT corresponds to the match output line of a single row or entry.

B. Inherent Voter Based Scheme

Almost all redundancy based strategies rely on a majority voting. The voter, therefore, becomes a critical unit for the correct operation of any TMR system. In this paper, we use a voterless fault- tolerant strategy to implement a robust TMR system design. With the inverter gate, performing TMR becomes very straightforward. All needed is to connect all of the different columns to interconnect through their inverters. Every time bits are sent from the different columns into a common interconnect, and the final value on the final inverter will be the majority of them. Therefore, there is no need for a centralized voter block. The inherent majority mechanism is a great benefit that allows us to remove the centralized voter unit of TMR system. A final inverter is used to convert the signal back to the original binary value.

In this work the TMR method is used to recover random errors. By simultaneously addressing three RAM columns and combining the result of the three, using the TMR approach, the errors on rows with one error can be recovered. Since TMR can't mask defects that appear as double or triple in the same row, these rows as named "Bad Rows". Also, because in triple possible columns, the number of defects and Bad Rows are different, these two factors are used to assign priority for applying TMR on every possible triple column. The algorithm is commenced by first applying TMR on triple possible columns that have less Bad Rows. If some triple possible columns have the same number of Bad Rows, the TMR is applied on ones that have more defects. If a triple column does not satisfy the requirements based on these factors, it is removed from the priority list. In the end, the final TMR configuration is applied to the most appropriate triple columns. The following pseudo code shows the triple column TMR identification procedure.

**Algorithm 2** finding sufficient triple columns for TMR

1: **while** TMR is sufficient **do**
2:  Count number of defects and Bad Rows in all possible triple columns
3:  Apply TMR on triple columns with fewer Bad rows. If triple possible columns have the same number of Bad Rows, apply TMR on triple columns with more defects
4:  Remove the recovered triple column from the fault matrix
5: **end while**

C. Nano RAM

Recent progress on nanodevice points to promising directions for future circuit design. However, nanofabrication techniques are not yet mature, making implementation of such circuits, at least on a large scale, in the near future infeasible. However, if photolithography could be used to implement circuits using these nanodevices, then hybrid nano/CMOS chips could be fabricated and the benefits of nanotechnology could be utilized immediately. A startup company, called Nantero, has developed and implemented a non-volatile nanotube random-access memory (NRAM) using photolithography.

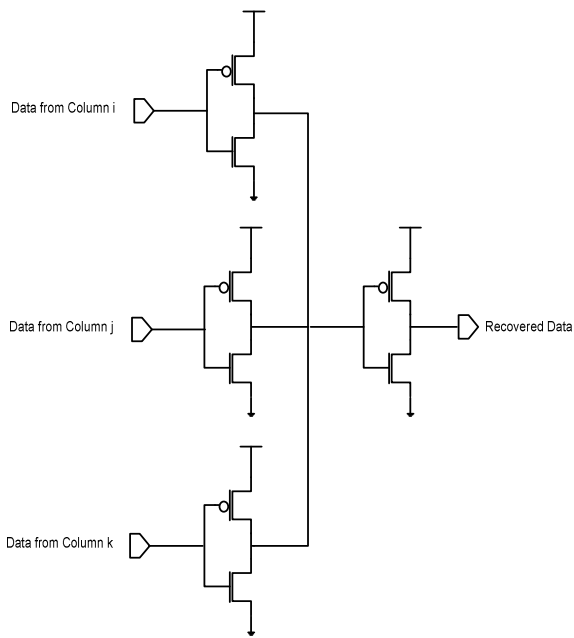

Figure 4. inherent voter used for the TMR

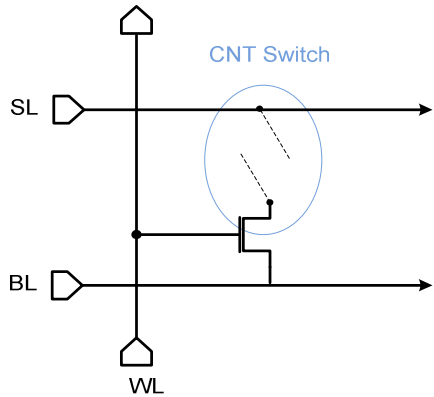

Figure 5. NRAM cell

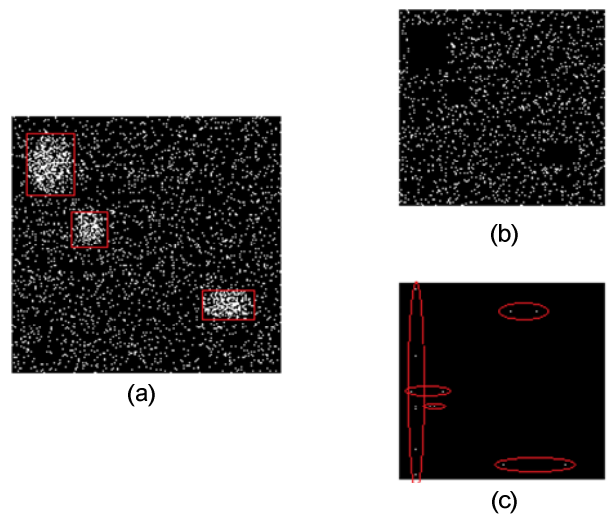

Figure 6. RAM (a) with uniform and Gaussian defects (b) after clusters repairing (c) after uniform and clusters repairing

As shown in Fig. 5 the CNT memory element is a 2-terminal variable resistance device, and is selected with a single NMOS transistor. The CNT fabric is located between two metal electrodes, which is defined and etched by photolithography, and forms the NRAM cell. When the CNTs are not in contact the resistance state of the fabric is high and represents a "0" state. When the CNTs are brought into contact, the resistance state of the fabric is low and represents a "1" state. In the "0" state, the CNTs are not in contact and remain in a separated state due to the stiffness of the CNTs resulting in a high resistance or low current measurement state between the top and bottom electrodes. In the "1" state, the CNTs are in contact and remain in a contacted state due to low resistance or high current measurement state between the top and bottom electrodes [11].

D.  RM-CAM and TMR based scheme

In this scheme the correlated defects are identified from random ones by Algorithm 1. The process is illustrated in Fig. 6 where the red boxes indicate RM-CAM regions which are removed from the fault matrix in Fig. 6(b). In the second phase, the remaining random defects are covered by TMR obtained from algorithm 2. Fig. 6(c) shows a rare situation where after the application of the TMR technique, some defects have remained and are shown by the red markers. Since after the application of the TMR method, most random defects are removed, in the final phase the defect clusters removed in phase 1 can be effectively mapped to healthy windows in the memory. It should be noted that the recovery of random errors using small window sizes in the RM-CAM requires a lot of entries in the RM-CAM and substantially increases the hardware complexity while the TMR provides a straightforward method to recover random errors. On the other hand the RM-CAM approach is more appropriate to remove a large window of errors. Therefore it is expected that the combination of the two techniques can be more effective in repairing actual nanodevice errors.

Fig.7 shows the entire mapping structure. If input addresses are located on clusters, two columns ROM is activated and give two vectors for row and column as placement vectors. These vectors add with primary row and column address. These addresses are mapped address as result of cluster mapping. If input addresses aren't located on cluster, outputs of two columns ROM must be "0". So, we use inverters for when don't any matches occur. For this reason, reverse of data in two columns ROM must be saved. Row mapped addresses enter to RAM decoder but column mapped address must be checked for TMR. For achieving this purpose we use a three columns ROM that give us three columns that must be voted.

III. EXPERIMENTAL RESULTS

MATLAB tools are used to carry out various simulations and analyses for evaluating the system reliability. The recovery rate is defined as the ratio of number of usable cells after repairing to the number of nondefective cells before repairing. It is assume that the number of cluster defects on the memory array is modeled by a Gaussian distribution and random defects by uniform distribution. In the simulations, the size of the memory array, initial mask size and cluster defects threshold is considered 256×256 bits, 25×25 bits and 150 respectively. The RM-CAM in this work is used for mapping defective windows to undefected regions and three columns ROM to map actual column addresses to triple column TMR locations.

While the exact manufacturing defect rate can vary from one process to another, defect rates as high as 10% have been reported [12]. For this reason, the proposed algorithm is tested with about 7% random

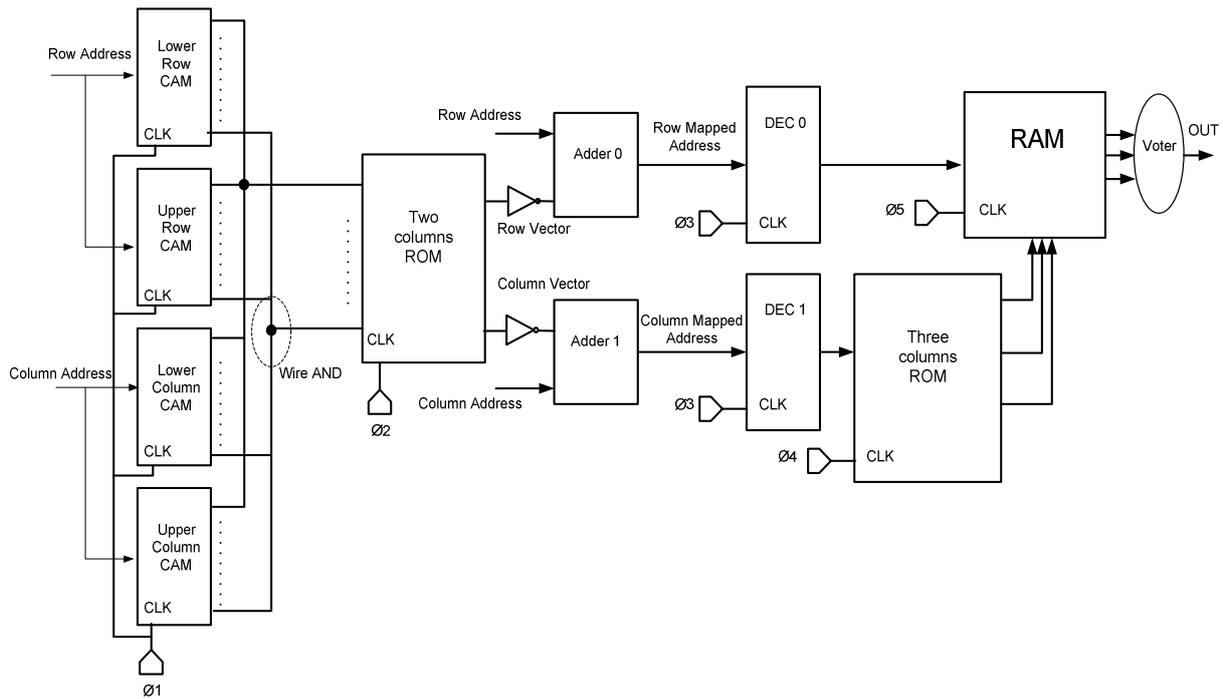

Figure 7. The completed cluster mapping and TMR structure

defects and 3% cluster defects atmost, resulting in a maximum total error rate of 10%.

In other methods with increases on uniform defect rate, RM-CAM size increases significantly but in the proposed method the RM-CAM size is constant. For n clusters, 4n-entries and for n TMR columns, n-entries of three columns ROM are necessary.

Fig. 8 shows the recovery rate for different uniform defect rates with an approximate 1.6% cluster defect rate. As it can be seen in the figure, initially with the increase of defect rate the recovery rate decreases rapidly. The reason can be refered to the fact that with more defects, the number of nondefective columns become less and thus the recovery rate decrease rapidly. The recovery rate drop continous until a defect rate of approximately 3%. From this point forward the recovery rate starts to increase slightly.

This observation can also be related to the fact that with the increase of defect rate, the number of TMR columns with three nondefective cells in same row become less, and thus less correct cells are left unused which results in the slight increase of recovery rate.

Fig. 9 illustrates the recovery rate under different cluster defect rates with a uniform random fault rate of approximately 7.5%. As the results show, the recovery rate decreases with the increase of cluster defect rate. The achieved result is because at higher defect rates and with more defect windows, the number of correct unused cells inside the maped window will increase and the recovery rate will drop.

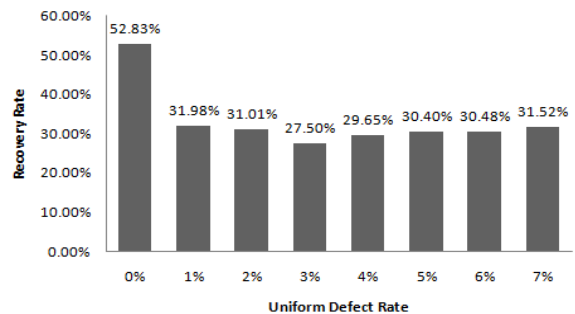

Figure 8. recovery rate with 1.5% to 1.8% cluster defects rate

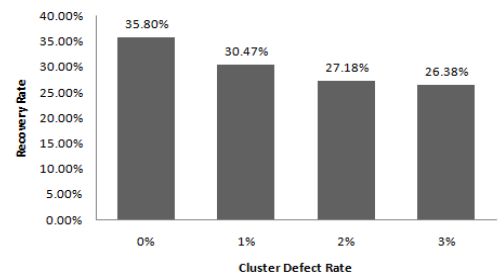

Figure 9. recovery rate with 7.4% to 7.6% uniform defects rate

## IV. CONCLUSIONS

In this paper, a combination method of TMR and RM-CAM was proposed to compactly store the defect map of a nanoscale memory. The presented algorithm identifies the local faults to be repaired by the RM-CAM mapping and also the random faults which should be compensated by the TMR technique. Current RAM reparing techniques require complex and subtantially complicated hardware to map and correct defects that are not clustered locally and are individually spreaded throught the memory with a random distibutuion. In the paper was shown that with a combinational technique is possible to use conventional table and mapping based approaches to rapir local and clustered defects while error tolerant based hardware which function best with random error can be used to repair random defects. The proposed technique shows better efficiency at high random defects compared with the previously reported methods.

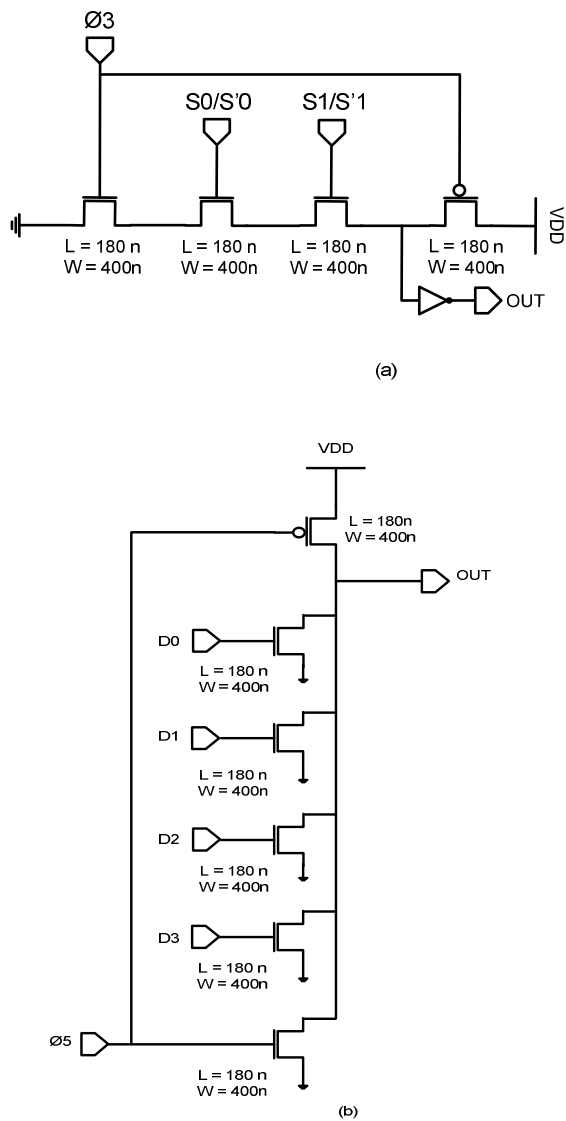

Figure 10. (a) decoder cell (b RAM cell

Decoder and RAM are modeled as shown in Fig.10.

Table I shows power, speed and number of transistors that needed for our structure for 32×32 and 64×64 RAM.

TABLE I. structure analysis

| RAM Size | Power (mW) | Speed (nS) | Number of Transistor |
|---|---|---|---|
| 32 × 32 | 0.006 | 5.6 | 6138 |
| 64 × 64 | 0.010 | 6.9 | 12356 |